\documentstyle[12pt]{article}
\newcommand{\be}{\begin{equation}}
\newcommand{\ee}{\end{equation}}
\newcommand{\ba}{\begin{eqnarray}}
\newcommand{\ea}{\end{eqnarray}}
\renewcommand{\thefootnote}{\fnsymbol{footnote}}
\begin{document}
\begin{center}
{\Large 
A new class of {\cal PT}-symmetric Hamiltonians with real
spectra.
}\\
\vspace{0.7cm}
{\large F.Cannata}$^{a,}$\footnote{{\it E-mail:}cannata@bo.infn.it},
{\large M.Ioffe}$^{b,}$\footnote{{\it E-mail:}ioffe@snoopy.phys.spbu.ru}, 
{\large R.Roychoudhury}$^{c,}$\footnote{{\it E-mail:}raj@isical.ac.in} and
{\large P.Roy}$^{d,}$\footnote{On leave of absence from
Indian Statistical Institute, Calcutta 700035,India;{\it
E-mail:}pinaki@isical.ac.in}\\
\vspace{0.2cm}
$^a${\small\it Dipartimento di Fisica and INFN, Via Irnerio 46,
40126 Bologna, Italy}\\
$^b${\small\it Department of Theoretical Physics, Institute of 
Physics, University of S.-Petersburg, Ulyanovskaya 1, S.-Petersburg 198904, 
Russia}\\
$^c${\small\it Physics and Applied Mathematics Unit, Indian 
Statistical Institute, Calcutta 700035, India}\\
$^d${\small\it Abdus Salam ICTP, Trieste, Italy}.
\end{center}
\vspace{0.5cm}
\noindent 
{\bf Abstract}
\bigskip

{\small
We investigate complex {\cal PT}-symmetric potentials,
associated with quasi-exactly solvable non-hermitian models involving
polynomials and a class of rational functions. 
We also look for special solutions
of intertwining relations of SUSY Quantum Mechanics providing a 
partnership between a real and a complex {\cal PT}-symmetric potential
of the kind mentioned above.
We investigate conditions sufficient to ensure the reality of the
full spectrum or, for the quasi-exactly solvable systems, the reality
of the energy of the finite number of levels.\\

\bigskip


} 
\bigskip
\setcounter{footnote}{0}
\renewcommand{\thefootnote}{\arabic{footnote}}
\section*{\large\bf 1.\quad Introduction}
\hspace*{3ex} 

While it is known that particular complex potentials 
have real spectra, irrespectively \cite{acdi}, \cite{junker},
\cite{khare} of
{\cal PT}-invariance, it is a widespread belief that {\cal PT}-symmetric
complex potentials have in general real spectra.
Within the general ongoing discussion of complex {\cal PT}-symmetric
potentials with real spectrum\footnote{Physical motivation
for this approach have been presented in \cite{bender}.} 
\cite{bender} (and references quoted therein) we study with special attention
quartic and sextic polynomial potentials and their generalizations
incorporating additional rational functions. In our discussion we will 
make use of algebraic techniques, of construction of quasi exactly 
solvable (QES) models \cite{turbiner}, \cite{ushveridze}, \cite{shifman} 
and methods of Supersymmetric Quantum Mechanics (SUSY QM) \cite{review},
\cite{review1}. 

We will not study exactly solvable non-hermitian models
as done for example in \cite{quesne}, \cite{levai} but  
our main goal is to show rigorous examples which are not
exactly solvable for which
a fixed amount of levels has real energies.
In the line
of the study of the sextic potentials in \cite{bagchi}
we investigate both quartic and sextic generalized potentials,
also including complex inverse square barriers \cite{znojil},
\cite{znojil1} \footnote{The {\cal PT}-symmetric harmonic
potential, including a complex barrier, has been investigated in 
\cite{znojil2}. For a recent discussion of such potential in the
contex of shape-invariance, cf. \cite{shape}. }. 
We provide also explicit constructions of models,
which again are not exactly solvable, but where the full spectrum of a 
non-hermitian potential is real, for example, due to 
SUSY isospectrality with a real potential \cite{acdi},
\cite{junker}, \cite{bagroy}.  
In Section 2 we discuss generalized quartic
potentials. In Section 3 the sextic case is analyzed.  

The main new results of this paper are:

- we have obtained a new class of quartic  complex {\cal PT}-symmetric
 potentials which are quasi-exactly solvable with a finite number
of levels with real energies;

- we have also discovered a generalized complex {\cal PT}-symmetric 
quartic potential with a real SUSY partner thus ensuring the reality
of the full spectrum;

- we found a generalized complex {\cal PT}-symmetric sextic potential
including an inverse square complex barrier with  solvable
ground state or ground and first excited states with a real
SUSY partner and thus with fully real spectrum.

\section*{\large\bf 2.\quad Quartic oscillator.}
\hspace*{3ex}
Within the general framework of complex potentials with real
spectrum we want to discuss some conditions sufficient to ensure
that a complex generalized polynomial potential of fourth order has indeed a
completely real spectrum. 
As outlined in the Introduction, we 
will try to use different approaches in order to
obtain complex potentials with real spectrum. 

The elementary algebraic approach, based on a complex
shift of coordinate, will apply
straightforwardly to a strict polynomial case.
More elaborated 
quasi-exactly solvable {\cal PT}-symmetric quartic potentials
have been investigated in \cite{boett} with special attention
to the reality of the QES-part (of finite dimensionality) of the spectrum,
here we find a much simpler quartic analogue of the celebrated
\cite{turbiner}, \cite{ushveridze}, \cite{shifman}, \cite{dunne}, 
QES sextic potential $V=x^6-(8j+3)x^2.$
Finally, based on the intertwining relations of SUSY QM
\cite{acdi}, \cite{junker}, \cite{bagroy},
we can demonstrate that the full spectrum of a modified quartic 
complex potential is real.

For a pure quartic case a sufficient condition for reality of
its spectrum
is provided if one can show that the complex potential can be 
reexpressed as a polynomial of $x_{\epsilon}\equiv (x+i\epsilon)$ 
with real coefficients, preserving {\cal PT}-invariance.
When this is the case the complexity of the potential is related only
to the complex shift of coordinate.

The general {\cal PT}-symmetric quartic complex 
potential reads: 
\be
V_4(x) = \rho x^4 + iax^3+bx^2+icx , \label{V4}
\ee
where $\rho =\pm 1$ and   
$a,b,c$ are, in general, arbitrary real parameters.
We study the restrictions on the values
of these parameters which are sufficient to ensure the real
character of the spectrum. The condition can be expressed in
a compact form as:
\be
V_4(x)=\rho x_{\epsilon}^4+Ax_{\epsilon}^2+B 
\label{f4}
\ee
with $A,B-$ real.
The appearance of only even powers follows from the
simultaneous requirement of {\cal PT}-invariance and of the
reality of the spectrum. 
The Eq.(\ref{f4}) leads to the constraints for Eq.(\ref{V4}) :
$$c=\frac{1}{8} a (a^2 + 4 \rho b)$$
and $\epsilon $ in Eq.(\ref{f4}) takes the value $\epsilon = \frac{1}{4} 
\rho a .$

We can now compare the potential (\ref{V4}) restricted by the condition
above with the quartic QES potential of \cite{boett} and find that
they are not compatible with quasi-exactly solvability condition\footnote{
With the notations of \cite{boett} (\ref{V4}) corresponds to the case
for $J=0.$}. 
Since the condition we have discussed
is only sufficient, this does not imply that the spectrum of \cite{boett}
can not be completely real.

Let us now consider pure quartic potentials in the framework of
SUSY QM \cite{review}, \cite{review1}. 
If one considers the simplest ansatz for the
superpotential $W=\rho x_{\epsilon}^2$ one can recognize 
that the ground state normalizability is lost for real values
of $\rho .$
Furthermore $\rho $ can not be anymore real if one requires 
{\cal PT}-invariance.
Indeed 
$$
W^2\pm W^{\prime}= \rho^2x_{\epsilon}^4\pm2\rho x_{\epsilon}
$$
is {\cal PT}-invariant only for pure imaginary $\rho$ and this is
inconsistent with Eq.(\ref{f4}).\footnote{One can easily check
that with imaginary $\rho$ one can have a normalizable ground
state wave function for one of the superpartners.}
We are thus unable to show that the full spectrum is real, but
we can at least show that a QES version (with $(2j+1)$ levels) 
of this model can be constructed following the approach of 
\cite{turbiner}, \cite{ushveridze}, \cite{shifman}.

We study the {\cal PT}-symmetric potential of the form:
\be
V(x)=-x_{\epsilon}^4-2iAx_{\epsilon}+Bj
\label{pot}
\ee
with $A, B$ - real constants and $j=0,1/2,1...$
It is useful to consider a peculiar "gauge" transformation of the 
wave functions:
\be
\Psi (x_{\epsilon})\equiv\exp (ix_{\epsilon}^3 / 3)\Phi (x_{\epsilon})
\label{ps}
\ee
for which the Schroedinger equation takes the form:
\be
[-\partial^2 - 2ix_{\epsilon}^2\partial - 2ix_{\epsilon}(A+1) 
+ Bj -E]\Phi (x_{\epsilon})=0; \qquad 
\partial\equiv\frac{d}{dx} .
\label{phi}
\ee
The corresponding non-hermitian operator (\ref{phi}) for $A=-(2j+1)$ 
can be written in terms of generators \cite{finkel} 
of the algebra $sl(2):$
\be
[-(J^{-})^{2} - 2iJ^{+} + Bj -E]\Phi (x_{\epsilon})=0;  
\label{H}
\ee

\be
J^- = \partial ;\quad J^{+} = x_{\epsilon}^{2} \partial -2j x_{\epsilon};
\quad J^{0} = x_{\epsilon} \partial - j;
\label{sl}
\ee
\be
[J^{0}, J^{\pm}] = 
\pm J^{\pm};\qquad [ J^{-}, J^{+} ] = 2J^{0}
\label{alg}
\ee
As usual \cite{turbiner}, \cite{ushveridze}, \cite{shifman}, 
polynomials $P_{2j+1}(x_{\epsilon})$
of degree $(2j+1)$ will provide a basis for the solutions 
$\Phi (x_{\epsilon})$ of the
Schroedinger equation (\ref{phi}). In our case the polynomials will be
complex and
will have coefficients which respect {\cal PT}-symmetry, i.e.
all the odd coefficients are imaginary when all the even are
real. One can explicitly check that all $(2j+1)$ lowest
eigenvalues are real. 

Within SUSY QM one can investigate, independently on the idea
of the shift $x\rightarrow x_{\epsilon},$ the partnership between
hermitian and non-hermitian Hamiltonians \cite{acdi}, \cite{junker},
\cite{bagroy}.
For reader's convenience we present the relevant condition \cite{acdi}
which expresses the above mentioned partnership with real
$V_-=W^2-W^{\prime}$ is given by:
\be
\exp{\int 2(Re W(x))dx}=\kappa (Im W(x)) , \label{acdi}
\ee
where $\kappa$ is an abitrary real integration constant.
{\cal PT}-invariance 
\footnote{The superpotential $W$ should
be odd under {\cal PT} transformation, like $\partial .$}
furthermore requires $(Re W(x))$ to be 
parity odd and $(Im W(x))$ to be even. 

We will consider polynomial superpotentials with addition
of the logarithmic derivative terms 
$P_M^{\prime}(x) / P_M(x)$ with lowest degrees of polynomials
$P_M \quad M=1,2,$
for which the integration
in (\ref{acdi}) is trivial. 
Guided by an ansatz discussed, for example, in \cite{fortsch},
we investigate the  
{\cal PT}-invariant
superpotential (with all constants $\beta , f, g$ - real):
\be
W(x)=ix^2 +i\beta + \frac{2fx}{1+fx^2} - \frac{ig}{1+igx}
\label{gener}
\ee
in order to construct two SUSY partner potentials of generalized
quartic class. We will pay attention to a suitable choice of
parameters which will make one partner real and the other partner
complex.
One partner potential:
\ba
V_-(x)=W^2(x)-W^{\prime}(x)=-x^4-2\beta x^2-\beta^2
+\frac{2}{g}+\frac{8f^2x^2}{(1+fx^2)^2}+\nonumber\\ 
\frac{1}{(1+fx^2)}
[4ix(-1+\beta f-\frac{f^2g}{f-g^2})
+\frac{2f(3g^2-f)}{f-g^2}]\nonumber\\
+\frac{2}{1+igx}[-\frac{1}{g} + \beta g-\frac{2fg^2}{f-g^2})],
\label{V-}
\ea
can be made real by the requirements:
\ba
-1 +\beta f-\frac{f^2g}{f-g^2} = 0;
\label{restr1}\\
-\frac{1}{g} + \beta g-\frac{2fg^2}{f-g^2}=0 .
\label{restr2}
\ea
The constant $f$ should be positive to avoid 
singularities of $V_- $ and the constraints (\ref{restr1}), (\ref{restr2})
can be solved for $f$ and $\beta$
in terms of the arbitrary negative\footnote{The constant $g$ must be negative
because it can be written as $g=-(fg)^2(f-g^2)^{-2} .$} constant $g .$ 
It is instructive to verify that equivalent constraints arise if
one calculates $\exp{\int 2(Re W(x))dx}$ and imposes that it should be
equal to $\kappa (Im W(x)).$

The other partner potential with the above substitutions remains complex:
\ba
V_+(x)=W^2(x)+W^{\prime}(x)=-x^4-2\beta x^2-\beta^2
+\frac{2}{g}+4ix+\nonumber\\
\frac{2f(f+g^2)}{(f-g^2)(1+fx^2)}
-\frac{2g^2}{(1+igx)^2}.
\label{V+}
\ea
Thus we realize an isospectrality between a complex $V_+(x)$ and
real $V_-(x)$ potential. 

The ground state of the superHamiltonian is the normalizable ground
state  wave function
\footnote{It is relevant to remark that this zero energy bound state of the 
real potential $V_-$ has a double 
degeneracy, corresponding to $\Psi_0^{(-)},\, \Psi_0^{(-)*},$ which are not
proportional to each other because of the $x$-dependent phase.}. 
associated to the real potential $V_- :$
\be
\Psi_0^{(-)}(x)= \exp{(-\int W(x)dx)} = 
\exp{\{i[\frac{\alpha x^3}{3}+\beta x]\}}
\frac{|1+igx|}{|1+fx^2|} .
\label{psi}
\ee
As a final comment to the above construction, we give a particular 
explicit solution of the constraints (\ref{restr1}), (\ref{restr2}):
$f=3g^2=-\beta^{-1}=2^{4/3}\cdot 3^{-1/3}.$ 
It is possible to generalize the quartic potential taking into
account additional terms of more complicate structure of
the type $P_M^{\prime}(x) / P_M(x)$  with $P_M$ a 
complex {\cal PT}-invariant polynomials. 

One can also try to include
an inverse square barrier by considering a {\cal PT}-odd superpotential:
\be
W(x)=ix_{\epsilon}^2-1/x_{\epsilon}
\label{Wbar}
\ee
leading to the pair of isospectral potentials:
\ba
V_-(x)=-x_{\epsilon}^4-4ix_{\epsilon}
\label{V-bar}\\
V_+(x)=-x_{\epsilon}^4+\frac{2}{x_{\epsilon}^2}
\label{V+bar}
\ea
The potential $V_+(x)$ is physically well defined
for $\epsilon\to 0$ on the half line where it is real. 
The most straightforward application
of this isospectrality between complex and real superpartners 
applies therefore for the so called radial problem \cite{acdi0}.

\section*{\large\bf 3.\quad Sextic potential}
\hspace*{3ex}
While the sextic potential has been studied thoroughly
both from algebraic and analytic points of view,
including {\cal PT}-symmetry \cite{turbiner}, \cite{dunne},
\cite{bagchi}, no systematic study 
in an algebraic framework
included an inverse square barrier. 
Here an attempt is made to accomplish this.
We discover that a rather general complex
sextic potential with a barrier of the form $2 / x_{\epsilon}^2$
has fully real spectrum though only $(2j+1)$ levels are
known analytically.
In order to match the discussion of the quartic potential of the
previous Section, let us start considering a complex 
{\cal PT}-symmetric potential:
\be
V_6(x) = \rho x^6 + 3iax^5+bx^4+icx^3+dx^2+iex ,
\label{V6}
\ee
where one can take $\rho =\pm$ and 
$a, b, c, d, e$ are again, in general, arbitrary real parameters.
The restrictions, sufficient to ensure the real
character of the spectrum, read:
\be
V_6(x)=\rho x_{\epsilon}^6+Ax_{\epsilon}^4+Bx_{\epsilon}^2+C 
\label{f6}
\ee
with $A,B,C$ - real.
Eq.(\ref{f6}) leads to the determination 
$\epsilon= \frac{1}{2} \rho a $ and to two constraints in Eq.(\ref{V6}) :
$$c=5 a^3 + 2 \rho ab $$
$$e=a^5+ \rho a^3b+\rho ad . $$
The additional constraint (see \cite{royroy})
\be
d=-\frac{75}{64}a^4+\frac{3}{8}a^2b+\frac{1}{4}b^2
-8j-3
\label{add}
\ee
leads to quasi-solvability of the model (\ref{V6}) for $(2j+1)$
levels with real energies.

While in the quartic case (\ref{pot}), discussed in the previous Section,
we could only ensure that the QES levels have real energies, for the
sextic case (\ref{V6}) we can build a QES model, which not only has
$(2j+1)$ levels with real energy, but has the full real spectrum.
The reason is that for the sextic potential the limit $\epsilon\to 0$
does not lead to any problem. 

For $j=0$ supersymmetrization is straightforward:
\be
W(x)=\rho^{1/2}x_{\epsilon}^3 + \frac{A}{2\rho^{1/2}}x_{\epsilon}
\label{susy}
\ee 
and the ground state wave function of the corresponding 
supersymmetrical hamiltonian takes the form:
\be
\Psi_0(x)=exp{(-\int W(x)dx)}=
exp{(- \frac{1}{4}\rho^{1/2}x_{\epsilon}^4-
\frac{A}{4\rho^{1/2}}x_{\epsilon}^2)} .
\label{wave}
\ee
This wave function is obviously normalizable for positive
values of $\rho ,$ taking the arithmetic determination of the root.
For larger values of $j$ quasi-solvability amounts
to add additional terms of the form $P_M^{\prime} / P_M$ in a superpotential
(\ref{susy}).

Methods based on Lie algebra \cite{turbiner}, \cite{ushveridze}, 
\cite{shifman}
to construct QES complex polynomial potentials
have been developed in \cite{royroy1}. 
Here we apply this method to a sextic (in terms of $x_{\epsilon}$) 
potential which can accomodate a $2 / (x_{\epsilon}^2)$
barrier\footnote{This construction can not be directly
generalized to a generic strength of the barrier with the same
structure for $W$ as in (\ref{susyb}).} \cite{znojil1}. 

Let us start with the potential
\be
V(x)=x_{\epsilon}^6+a_4x_{\epsilon}^4+a_2x_{\epsilon}^2+
\frac{a_{-2}}{x_{\epsilon}^2} 
\label{invers}
\ee
where $a_k$ are real.
It is convenient to define the superpotential
\be
W(x_{\epsilon}) = x_{\epsilon}^3 + ax_{\epsilon}+\frac{\gamma}{x_{\epsilon}}
\label{susyb}
\ee 
with real values for the constants $a,\gamma.$
After the "gauge" transformation
\be
\Psi (x_{\epsilon})\equiv \exp (-\int{W(x_{\epsilon})dx})\phi (x_{\epsilon})
\label{pphi}
\ee
the Schroedinger equation takes the form:
\be
[-\partial^2 + 2W(x_{\epsilon})\partial + (V(x_{\epsilon})-W^2(x_{\epsilon})
+W^{\prime}(x_{\epsilon})-E)]\phi (x_{\epsilon})=0 .
\label{newsch}
\ee
By the substitutions $z\equiv x_{\epsilon}^2$ and $\chi (z)\equiv 
\phi (x_{\epsilon})$ and by the requirement:
\be
a=a_4/2;\quad \gamma (\gamma+1)=a_{-2} , 
\label{const}
\ee 
it can be further transformed to:
\be
\{ -4z\frac{d^2}{dz^2}+(4z^2+4az+4\gamma -2)\frac{d}{dz} +
(a_2+3-a^2-2\gamma )z +a-2a\gamma -E \}\chi (z)=0 .
\label{abc}
\ee
Introducing the same $sl(2)$ generators as in the previous Section
\be
J^- = \frac{d}{dz} ;\quad J^{+} = z^{2} \frac{d}{dz} -2j z;
\quad J^{0} = z \frac{d}{dz} - j ,
\label{slz}
\ee
which satisfy (\ref{alg}), the Schroedinger equation (\ref{abc}) 
finally can be cast in algebraic form as:
\be
(-4J^0J^{-} +4J^{+}+(4\gamma -2-4j)J^-+4aJ^0 + \bar{E})\chi (z)=0 , \label{Hz}
\ee
where $\bar{E}=a-E-2a\gamma +4aj $ 
and 
$$a_2=a_4^2/4-8j-3+2\gamma .$$
Thus the potential of (\ref{invers}) becomes:
\be
V(x)=x_{\epsilon}^6+2ax_{\epsilon}^4+(a^2-8j-3+2\gamma )x_{\epsilon}^2+
\frac{\gamma (\gamma +1)}{x_{\epsilon}^2} 
\label{form}
\ee
and is QES for $j=0,1/2,1,...$
The corresponding wave functions can be written as:
\be
\Psi (x_{\epsilon})= P_{2j+1}(x_{\epsilon}) \exp (-\int W(x_{\epsilon})dx) .
\label{pppsi}
\ee
Notice that for $j=0$ the potential $$V=V_-+a(1-2\gamma )
=W^2-W^{\prime}+a(1-2\gamma )$$ 
has the ground state
\be
\Psi_0 (x_{\epsilon})=\exp (-\int W(x_{\epsilon})dx).
\label{pppsii}
\ee
with energy given by $E_0=a(1-2\gamma ).$
Taking $\gamma = +1,$ the partner potential $V_+=W^2+W^{\prime},$
has no barrier term and is a polynomial with real coefficients.
Thus the spectrum of (\ref{form}) for $j=0$ is real though only
one level is known analytically. 

For $j=1/2$ one can also obtain the reality of the
full spectrum using SUSY techniques \cite{fortsch}
as follows. Starting from
\be
W(x) = x_{\epsilon}^3 + ax_{\epsilon}+ \frac{\gamma}{x_{\epsilon}}
-\frac{2fx_{\epsilon}}{1+fx_{\epsilon}^2} .
\label{last}
\ee
it is easy to show that $W^2-W^{\prime}$ coincides with potential
(\ref{form}) for $j=1/2$ and $\gamma =1$ apart from a constant 
for $f=a+\sqrt{a^2-2}.$ If 
$a>\sqrt{2},$ the partner potential
$W^2+W^{\prime}$
 has no barrier term and,
having a well defined
\footnote{$f$ is positive.} 
behaviour for $\epsilon\to 0,$ has real
spectrum.

\section*{\large\bf\quad 4. Conclusions.}
\hspace*{3ex}
In this paper we have discovered a new class of {\cal PT}-symmetric
quartic and sextic potentials with real energy spectrum. We have found
a class of quartic potentials which have a real superpartner,
thus ensuring real spectrum for them. 
A {\cal PT}-symmetric sextic potential with a $2 / x_{\epsilon}^2$
barrier term can be accomodated in our scheme of quasi-exactly solvable
{\cal PT}-symmetric potentials with a full real spectrum.
As an outlook on future perspectives we would like to point out
that it is possible to formulate an innovated \cite{znojil} SUSY framework
where the two superpartners live in two different  half planes
of the complex $x-$plane achieving thereby normalizability
of both superpartner wave functions. 
This approach should
be particularly interesting for the quartic type
Eq.(\ref{psi})) because it 
would enlarge the class of {\cal PT}-symmetric
potentials with real spectrum. 
An interesting feature of this approach will be the
violation of the Witten criterion similarly to higher order 
SUSY QM \cite{acdi0}.

\section*{\large\bf\quad Acknowledgments.}
\hspace*{3ex}
Two of the authors (M.I. and R.R.) thank INFN and University of Bologna 
for warm hospitality. The work of M.I. was partially supported by RFBR 
(No. 99-01-00736). P.R. would like to thank the Abdus Salam ICTP 
for hospitality.

\end{document}